\documentclass{article}
\usepackage[T2A]{fontenc}
\usepackage[utf8]{inputenc}
\usepackage[english,russian]{babel}
\usepackage[tbtags]{amsmath}
\usepackage{amsfonts,amssymb,mathrsfs,amscd}
\usepackage{graphicx}
\usepackage{epstopdf}
\usepackage[hyper]{msb-a}

\makeatletter
\gdef\No{{\select@language{russian}\textnumero}}
\makeatother

\JournalName{}
\numberwithin{equation}{section}
\theoremstyle{plain}
\newtheorem{theorem}{Теорема}
\newtheorem{lemmaa}{Лемма}[section]

\theoremstyle{definition}

\newtheorem{proof}{Доказательство}

\begin{document}

\title{Эффективный поиск минимального дерева на точках пространства в $l_1$-норме}
\author[K.\,V.~Kaymakov]{К.\,В.~Каймаков}
\address{Национальный исследовательский университет <<Высшая школа экономики>>, Покровский б-р, 11с4, Москва, 109028, Россия}
\email{kirill.kaymakov@mail.ru}

\author[D.\,S.~Malyshev]{Д.\,С.~Малышев}
\address{Национальный исследовательский университет <<Высшая школа экономики>>, ул. Большая Печёрская, 25/12, 603155, Нижний Новгород, Россия;\\
Московский физико-технический институт, Институтский переулок, 9, 141700, Московская область, г. Долгопрудный, Россия}
\email{dsmalyshev@rambler.ru}

\date{}
\udk{519.163}

\maketitle

\begin{fulltext}

\begin{abstract}
В данной работе рассматривается задача о минимальном остовном дереве (кратко, ЗМОД) на
произвольном множестве $n$ точек $d$-мерного пространства в $l_1$-норме.
Для этой задачи при каждом фиксированном $d\geq 2$ известен алгоритм сложности $O\big(n\cdot (\log\,n + \log^{r_d}\,n\cdot \log\log\,n)\big)$,
где $r_d\in \{0,1,2,4\}$ при $d\in \{2,3,4,5\}$ и $r_d=d$ при $d\geq 6$. Для $d=3$ известно
улучшение этого результата до сложности $O(n\cdot \log\,n)$. В настоящей работе при любом
фиксированном $d\geq 2$ для решения рассматриваемой ЗМОД предлагается алгоритм со сложностью $O(n\cdot \log^{d-1}\,n)$,
что улучшает предыдущее достижение при $d\geq 5$.

Библиография: 23 названий.
\end{abstract}

\address{НИУ ВШЭ, ООО <<Коулмэн Тех>>, МФТИ}

\markright{Эффективный поиск...}


\section{Введение}

\emph{Задача о минимальном остовном дереве} (кратко, ЗМОД)~---~это классическая задача комбинаторной
оптимизации. Эта задача для заданного связного взвешенного по ребрам графа состоит в том, чтобы
найти в нем \emph{остовное дерево}, т.е. связный ациклический подграф, содержащий все вершины графа, с минимальной суммой весов его ребер.
ЗМОД имеет многочисленные приложения не только в алгоритмической теории графов (см., например, обзор \cite{GH85}), но еще и в анализе данных \cite{ABKY88}, компьютерном зрении \cite{FH04}, проектировании интегральных схем \cite{O04}, анализе финансовых рынков \cite{M99} и многих других.

К настоящему времени для решения ЗМОД на графах c $n$ вершинами и $m$ ребрами разработано несколько эффективных алгоритмов:
алгоритм Борувки со сложностью $O(m\cdot\log\,n)$ (см. \cite{B26,T83}), алгоритм Краскала со сложностью\\ $O(m\cdot \log\, n)$ (см. \cite{T83,K56}), алгоритм Прима
со сложностью $O(m+n\cdot\log\,n)$ (см. \cite{P57,FT87}), алгоритм Шазелла со сложностью $O\big(m\cdot Ack^{-1}(m,n)\big)$ \cite{Ch00}, где $Ack^{-1}(\cdot,\cdot)$~--- обратная функция Аккермана, и ряд других. Все эти алгоритмы являются детерминированными.

На настоящее время открытой проблемой является разработка детерминированного  алгоритма решения ЗМОД со сложностью $O(m)$ (или доказательство его отсутствия), и на настоящее время алгоритм Шазелла является рекордным по вычислительной сложности среди детерминированных алгоритмов. Ожидаемой линейной сложности удается добиться за счет рандомизации, соответствующий алгоритм был разработан Каржером, Клейном и Тарджаном в \cite{KKT95}. Для некоторых отдельных классов графов
или семейств классов графов удается построить детерминированные алгоритмы решения ЗМОД линейной сложности. Например, для классов графов, отличных от множества всех графов и замкнутых относительно удаления вершин и ребер, а также стягивания ребер, ЗМОД может быть решена за линейное время \cite{M02}.

Вызывает особый интерес класс \emph{геометрических ЗМОД}, т.е. задач вычисления МОД в полных графах на системах $n$ точек $d$-мерного пространства с \emph{$l_p$-нормой} $||x,y||_p=\sum\limits_{i=1}^{d}|x_i-y_i|^p$, где $p\geq 1$, между точками. Во-первых, геометрические ЗМОД возникают
в алгоритмах кластеризации (см., например, \cite{ABKY88}) и эффективные алгоритмы решения этих ЗМОД ускоряют сами алгоритмы кластеризации. Во-вторых, некоторые из них могут быть решены за время, меньшее, чем $O(m)=O(n^2)$. Например, при $d=2$ и $p=1$ за время $O(n\cdot \log\,n)$ \cite{GS83,ZSN02}, при $d=2$ и $p=2$ за время $O(n\cdot \log\,n)$ \cite{SH75}, для $p=2$ за время $O\big((n\cdot\log\,n)^{\frac{4}{3}}\big)\,(d=3)$ и для любого $\epsilon>0$ за время $O(n^{2-\frac{2}{\lceil\frac{d}{2}\rceil1}+\epsilon})\,(d\geq 4)$ \cite{AESW91}. Для $d=p=2$ соответствующая геометрическая ЗМОД может быть решена (см. работу \cite{D92})
рандомизированным алгоритмом с ожидаемым временем работы $O(n\cdot \log_{*}\,n)$, где $\log_{*}\,n$ --- итерированный логарифм.

Для случая $p=1$ наиболее важные результаты представлены в работах \cite{GBT84} и \cite{KLN97}. В первой из них доказано,
что при любом фиксированном $d\geq 2$ соответствующая геометрическая ЗМОД может быть решена за время $$O\big(n\cdot (\log\,n + \log^{r_d}\,n\cdot \log\log\,n)\big),$$ где $r_d\in \{0,1,2,4\}$ при $d\in \{2,3,4,5\}$ и $r_d=d$ при $d\geq 6$. Во второй показано, что для $d=3$ она может быть решена со сложностью $O(n\cdot \log\,n)$.

В настоящей работе при $d\geq 5$ улучшается результат из работы \cite{GBT84}. А именно, следующее утверждение составляет основной результат нашей работы:

\begin{theorem}
\label{t1}
Для любого фиксированного $d\geq 2$ задача о минимальном остовном дереве на любой системе $n$ точек пространства $\mathbb{R}^d$ c $l_1$-нормой между точками может быть решена за время $O(n\cdot \log^{d-1}\,n)$.
\end{theorem}

\section{Свойство близости и его значение}

В основе многих алгоритмов решения геометрических ЗМОД лежит следующая идея~--- каким-нибудь образом построить разреженный надграф МОД
заданных точек, а затем вызвать эффективный алгоритм решения полученной разреженной ЗМОД, например, алгоритм Шазелла или алгоритм Прима. Время работы такого рода алгоритма есть $$O\big(T(n,m) + m\cdot Ack^{-1}(m,n)\big)\,\,\text{или}\,\,O(T(n,m) + m + n\cdot\log\,n),$$
где $T(n,m)$ --- время построения надграфа МОД. Способ построения данных надграфов был предложен в работе \cite{Y82}, причем для $p=2$ в качестве этого надграфа может выступать триангуляция Делоне (см., например, \cite{AESW91,D92}).

Всюду далее точки $\mathtt{x}$ пространства $\mathbb{R}^d$ и их координаты $x_i$ связываются так:
$\texttt{x}=(x_1,\ldots,x_d)$. Способ из \cite{Y82} основан на покрытии (относительно заданных точки $\texttt{s}$ и нормы $||\cdot,\cdot||$) пространства $\mathbb{R}^d$
областями $R_1,\ldots,R_{k(\texttt{s})}$, где:
\begin{itemize}
\item $\texttt{s}$ принадлежит всем этим областям,
\item каждая из областей $R_i$ обладает \emph{свойством близости} (\emph{относительно} $\texttt{s}$) --- для любых точек $\texttt{p},\texttt{q}\in R_i$ выполнено $||\texttt{p},\texttt{q}||\leq \max\big(||\texttt{s},\texttt{p}||,||\texttt{s},\texttt{q}||\big)$.
\end{itemize}

В \cite{Y82} (см. Теорему 3.2 той работы) было показано, что для любого конечного множества $V$ точек совокупность всех ребер вида $\texttt{s}\texttt{s}_i$, где $\texttt{s}\in V$ и $\texttt{s}_i$ --- ближайший по $||\cdot,\cdot||$ сосед $\texttt{s}$ в $R_i\setminus \{\texttt{s}\}$, порождает надграф некоторого оптимального решения ЗМОД на $V$ с не более чем $\sum\limits_{\texttt{s}\in V} k(\texttt{s})$ ребрами. Конкретные разбиения со свойством близости рассматривались и применялись в работах \cite{ZSN02,GBT84,Y82}.

Для любых $p\geq 1$ и точки $\texttt{s}$ в работе \cite{Y82} (см. Леммы 4.2 -- 4.4, 7.1 -- 7.2 той работы, а также ассоциированное алгоритмическое обеспечение) представлено покрытие $\Upsilon_{d,p}(\texttt{s})$ пространства ${\mathbb R}^d$, которое

\begin{itemize}
\item состоит только из \emph{$d$-симплексов} --- многогранных конусов с центром в $\texttt{s}$, образованных ровно $d$ линейными неравенствами,
\item обладает свойством близости относительно $\texttt{s}$ и $l_p$-нормы.
\end{itemize}
Опишем идею построения $\Upsilon_{d,p}(\texttt{s})$ из \cite{Y82}. Через $B_d$ обозначается множество $\{+1,-1\}^d$, а через $[r]$ обозначается множество $\{1,\ldots,r\}$. Для вектора $\alpha \in B_d$ и точки $\texttt{s}$ через $Ort_{d,\alpha}(\texttt{s})$
обозначается ортант

$$\{\texttt{x}:\, \alpha_i\cdot (x_i-s_i)\geq 0\,\, \forall i\in [d]\}.$$
Для любого $\alpha\in B_d$, начиная с $Ort_{d,\alpha}(\texttt{s})$, каждый текущий $d$-симплекс $R$ с центром
в $\texttt{s}$ и образующими $\texttt{e}_{1,R},\ldots,\texttt{e}_{d,R}$ подразбивается своим барицентрическим направлением
$\frac{\sum\limits_{i=1}^{d} \texttt{e}_{i,R}}{d}$ на более узкие $d$-симплексы. Процесс повторяется до тех пор, пока
каждый из конусов не станет достаточно узким. Более точно, критерием остановки (см. Лемму 4.3 из \cite{Y82}) процесса подразбиения является выполнение неравенства
$$\max\limits_{i,j\in [d]} \angle(\texttt{e}_{i,R},\texttt{e}_{j,R})<\arcsin(\frac{1}{2d^\frac{3}{2}}).$$

Положим $\Upsilon_{d}(\texttt{s})=\Upsilon_{d,1}(\texttt{s})$. Очевидно, что каждый элемент $\Upsilon_{d}(\texttt{s})$
описывается системой неравенств вида $\textmd{A}\cdot(\texttt{x}-\texttt{s})^{T}\geq \texttt{0}$, где каждое $a_{ij}$ зависит
только от $d$ и $\alpha$, а $\texttt{0}$ --- начало координат. Элементы $\Upsilon_{d}(\texttt{s})$, составляющие ортант $Ort_{d,\alpha}(\texttt{s})$,
будем обозначать через $R_{d,\alpha,1}(\texttt{s}),R_{d,\alpha,2}(\texttt{s}),\ldots R_{d,\alpha,k_d}(\texttt{s})$, а матрицу системы неравенств в описании $R_{d,\alpha,i}(\texttt{s})$ будем обозначать через $\textmd{A}_{d,\alpha,i}$ для любого $i\in [k_d]$. Считаем, что $R_{d,\alpha,i}(\texttt{s})$ и $R_{d,-\alpha,i}(\texttt{s})$ являются симметричными относительно $\texttt{s}$ для любых $\alpha \in B_d, i\in [k_d]$.

\section{Описание предлагаемого алгоритма, обоснование его корректности и трудоемкости}

\subsection{Метод сканирующей гиперплоскости и его применение}

\emph{Метод сканирующей гиперплоскости}~---~это способ обработки заданной системы точек пространства,
состоящий в их упорядочивании по какому-нибудь критерию (или каких-нибудь событий, ассоциированных с точками) и последующем проходе по ним.
Например, в известном алгоритме Бентли-Оттмана \cite{BO79} для поиска всех пар пересекающихся среди заданных отрезков на плоскости такими событиями являются
<<начало отрезка>>, <<конец отрезка>> и <<пересечение отрезков>>. Это позволяет найти все пары
пересекающихся отрезков среди $n$ заданных на плоскости за время $O(n\cdot \log\,n+k)$, где $k$~---~количество пар пересекающихся отрезков, что лучше при не очень больших $k$, чем наивный переборный алгоритм сложности $\Theta(n^2)$.

Для плоского случая метод сканирующей прямой для построения надграфа МОД использовался в работе \cite{ZSN02}.
В этой работе мы обобщаем этот прием для произвольного $d\geq 3$. А именно, заданное множество точек мы будем упорядочивать по следующим линейным критериям $\preceq_{\alpha}$, где $\alpha\in B_d$:
$$\texttt{x}=(x_1,\ldots,x_d)\preceq_{\alpha} \texttt{y}=(y_1,\ldots,y_d)\Longleftrightarrow \sum\limits_{i=1}^{d}\alpha_i\cdot x_i \leq \sum\limits_{i=1}^{d} \alpha_i\cdot y_i.$$

Нам понадобится следующее утверждение:

\begin{lemmaa}
\label{l1}
Для любых $\alpha\in B_d$ и точек $\texttt{x},\texttt{y},\texttt{s}\in \mathbb{R}^d$ выполнено:

$$\texttt{x},\texttt{y}\in Ort_{d,\alpha}(\texttt{s}),\texttt{x}\preceq_{\alpha}\texttt{y}\Longrightarrow||\texttt{y},\texttt{s}||_1\geq ||\texttt{x},\texttt{s}||_1.$$
\end{lemmaa}

\proof Имеем

$$||\texttt{y},\texttt{s}||_1-||\texttt{x},\texttt{s}||_1=\sum\limits_{i=1}^{d}\big(|y_i-s_i|-|x_i-s_i|\big)=\sum\limits_{i=1}^{d}\big(|\alpha_i\cdot (y_i-s_i)|-|\alpha_i\cdot (x_i-s_i)|\big)=$$
$$=\sum\limits_{i=1}^{d}\big(\alpha_i\cdot (y_i-s_i)-\alpha_i\cdot (x_i-s_i)\big)=\sum\limits_{i=1}^{d}\big(\alpha_i\cdot (y_i-x_i)\big)\geq 0,$$
т.к. в условиях Леммы \ref{l1} выполнено $$|z-s_i|=|\alpha_i\cdot (z-s_i)|=\alpha_i\cdot (z-s_i)\geq 0\,\,\forall z\in \{x_i,y_i\},\, \forall i\in [d].\,\blacksquare$$

Нетрудно видеть, что $\texttt{y}\in R_{d,\alpha,i}(\texttt{x}) \Longleftrightarrow \texttt{x}\in R_{d,-\alpha,i}(\texttt{y})$. Отсюда и из Леммы \ref{l1}
следует, что если $\texttt{x}\in R_{d,-\alpha,i}(\texttt{y}),\texttt{x}\in R_{d,-\alpha,i}(\texttt{z})$ и $\texttt{y} \preceq_{\alpha} \texttt{z}$, то
$||\texttt{y},\texttt{x}||_1\leq ||\texttt{z},\texttt{x}||_1$. Тем самым, для любых $d,\alpha,i$ и точки $\texttt{x}$ первая справа по порядку $\preceq_{\alpha}$ точка $\texttt{y}\neq \texttt{x}$, для которой $\textmd{A}_{d,\alpha,i}\cdot(\texttt{y}-\texttt{x})^{T}\geq \texttt{0}$, будет ближайшим к $\texttt{x}$ элементом множества $R_{d,\alpha,i}(\texttt{x})\setminus \{\texttt{x}\}$.

\subsection{Эффективный динамический интервальный поиск и его применение}

\emph{Ортогональный интервальный поиск}~---~это поиск в заданном множестве $n$ точек $P\subset \mathbb{R}^d$ для заданных $\texttt{x},\texttt{y}\in (\mathbb{R}\cup \{-\infty,+\infty\})^d$, где для любого $i\in [d]$ выполнено $x_i\leq y_i$, подмножества $$L=\big\{\texttt{z}\in P:\,\text{для любого}\,\,i\in [d]\,\,\text{выполнено}\,\,x_i\leq z_i\leq y_i\big\}.$$ При этом $P$ может быть как статическим, так и динамическим, т.е. допускать вставки и/или удаления точек. В работе \cite{MS90} (см. Теоремы 8 и 9 из той работы) было показано, что для $d\geq 2$ при помощи специальной версии динамического дерева отрезков множество $L$ может быть вычислено за время

$$O\big(\log^{d-1}\,n\cdot \log\log\,n+|L|\big).$$
При этом время построения этого дерева есть

$$O(n\cdot \log^{d-1}\,n\cdot \log\log\,n)$$
при используемой памяти $O(n\cdot \log^{d-1}\,n)$, а время обновления (вставки и удаления элемента) есть  $$O(\log^{d-1}\,n\cdot \log\log\,n).$$
Отметим, что если поддерживаются только вставки или только удаления точек, то сомножитель $\log\,\log\,n$ в оценках сложности может быть опущен (см. пункт e
Теорем 8 и 9 из работы \cite{MS90}).

Для любых $d,\alpha,i$ и точки $\texttt{s}$ принадлежность точек из $P$ конусу $R_{d,\alpha,i}(\texttt{s})$ определяется ортогональным интервальным поиском.
Напомним, что $$R_{d,\alpha,i}(\texttt{s})=\{\texttt{x}:\,\, \textmd{A}_{d,\alpha,i}\cdot \texttt{x}^{T}\geq \textmd{A}_{d,\alpha,i}\cdot \texttt{s}^{T}\}$$
и что $\textmd{A}_{d,\alpha,i}$ имеет ровно $d$ строк.
Вычислим

$$(\texttt{s}')^T=\textmd{A}_{d,\alpha,i}\cdot \texttt{s}^{T}\,\,\text{и}\,\,P'=\big\{(\texttt{x}')^T=\textmd{A}_{d,\alpha,i}\cdot \texttt{x}^{T}:\,\,\texttt{x}\in P\big\}.$$
Заметим, что
$$P_{d,\alpha,i}(\texttt{s})=R_{d,\alpha,i}(\texttt{s})\cap P=\{\texttt{x}'\in P':\,\, s'_i \leq x'_i\,\,\, \forall i\in [d]\}.$$

Тем самым, предобработка $P'$ (т.е. построение соответствующего динамического дерева отрезков) занимает $O(n\cdot \log^{d-1}\,n)$
время, подмножество $P_{d,\alpha,i}(\texttt{s})$ может быть перечислено за время

$$O\big(|P_{d,\alpha,i}(\texttt{s})|+\log^{d-1}\,n\big),$$
удалено из $P$ за время

$$O\big(|P_{d,\alpha,i}(\texttt{s})|\cdot \log^{d-1}\,n\big).$$

\subsection{Формальное описание алгоритма}

Для множества $n$ точек $P\subset \mathbb{R}^d$ наш алгоритм может быть описан следующим образом:
\medskip

\emph{Шаг} 0. Для всех $\alpha\in B_d$ и $i\in [k_d]$ найти матрицы $\textmd{A}_{d,\alpha,i}$.
\smallskip

\emph{Шаг} 1. В двойном цикле по $\alpha\in B_d$ и $i\in [k_d]$ выполнить:
\smallskip

\hspace{0.2cm}\emph{Шаг} 1.1. Вычислить $P':=\{\texttt{x}'=\textmd{A}_{d,-\alpha,i}\cdot \texttt{x}^{T}:\,\,\texttt{x}\in P\}$ и
 построить

\hspace{1.8cm}динамическое дерево отрезков для $P'$.
\smallskip

\hspace{0.2cm}\emph{Шаг} 1.2. Упорядочить точки из $P$ по критерию $\preceq_{\alpha}$.
\smallskip

\hspace{0.1cm} \emph{Шаг} 1.3. Итерируясь по $\texttt{s}\in P$ в соответствии с порядком $\preceq_{\alpha}$:
\smallskip

\hspace{0.4cm}\emph{Шаг} 1.3.1. Найти $(\texttt{s}')^T:=\textmd{A}_{d,-\alpha,i}\cdot \texttt{s}^{T}$ и $P(\texttt{s}):=\{\texttt{x}'\in P'\setminus \{s\}: s'_i \leq x'_i\,\, \forall i\}$

\hspace{2.2cm}и ближайший к $\texttt{s}$ элемент $\texttt{s}''\in P(\texttt{s})$.
\smallskip

\hspace{0.4cm}\emph{Шаг} 1.3.2. Соединить ребром $\texttt{s}$ и $\texttt{s}''$.
\smallskip

\hspace{0.4cm}\emph{Шаг} 1.3.3. Выполнить $P':=P'\setminus P(\texttt{s})$.
\smallskip

\emph{Шаг} 2. К построенному на Шаге 1 графу применить алгоритм Шазелла или алгоритм Прима.
\medskip

\subsection{Доказательство корректности и вывод оценки трудоемкости}

Докажем корректность нашего алгоритма. Для любых $\alpha\in B_d,i\in [k_d],\texttt{s}\in P$
он рассматривает конус $R_{d,-\alpha,i}(\texttt{s})$ и точку $\texttt{s}''$. Эта точка $\texttt{s}''$
принадлежит $R_{d,-\alpha,i}(\texttt{s})$, что означает $\texttt{s}'' \preceq_{\alpha} \texttt{s}$.
Нетрудно видеть, что $\texttt{s}$ --- первая относительно порядка $\preceq_{\alpha}$ точка, принадлежащая
$R_{d,\alpha,i}(\texttt{s}'')$. По Лемме \ref{l1} это означает, что $\texttt{s}''$ --- ближайший к $\texttt{s}$
элемент конуса $R_{d,-\alpha,i}(\texttt{s})\setminus \{\texttt{s}\}$. По Теореме 3.2 из \cite{Y82} построенное множество
ребер образует надграф некоторого МОД. Это доказывает корректность нашего алгоритма.

Оценим вычислительную сложность нашего алгоритма. Время вычисления всех матриц $\textmd{A}_{d,\alpha,i}$ (т.е. сложность Шага 0) не зависит от $n$.
Количество итераций двойного внешнего цикла Шага 1 также не зависит от $n$. Время выполнения каждого из Шагов 1.1 есть
$$O(d^2\cdot n + n\cdot \log^{d-1}\,n),$$
а время выполнения каждого из Шагов 1.2 есть $O(d\cdot n + n\cdot \log\,n)$. Время выполнения каждого Шага 1.3.1 есть
$$O(d^2 + |P(\texttt{s})|+\log^{d-1}\,n),$$
время выполнения каждого Шага 1.3.2 есть $O(1)$, а время выполнения каждого Шага 1.3.3 есть $$O\big(|P(\texttt{s})|\cdot \log^{d-1}\big).$$
Нетрудно видеть, что если $\texttt{s}_1 \preceq_{\alpha} \texttt{s}_2$, то $P(\texttt{s}_1)\cap P(\texttt{s}_2)=\emptyset$. Поэтому $\sum\limits_{\texttt{s}\in P} |P(\texttt{s})|\leq n$ и общее время выполнения цикла из Шага 1.3 при фиксированном $d$ есть $O(n\cdot \log^{d-1}\,n)$. Тем самым, при фиксированных $d,\alpha,i$ общая сложность всех итераций вложенного цикла из Шага 1 есть $O(n\cdot \log^{d-1}\,n)$.

Нетрудно видеть, что количество ребер $m$ получившегося графа удовлетворяет неравенству $m\leq 2^d\cdot k_d\cdot n$. Поэтому сложность Шага 2 при использовании алгоритма Прима есть $O(2^d\cdot k_d\cdot n + n\cdot \log\,n)$. Тем самым, общая сложность нашего алгоритма при фиксированном $d$ есть $O(n\cdot \log^{d-1}\,n)$.

Таким образом, доказана Теорема \ref{t1}.

\section{Заключение}

В настоящей работе рассматривалась задача о минимальном остовном дереве (ЗМОД) для точек пространства $\mathbb{R}^d$
и $l_1$-расстояния между парами точек. ЗМОД имеет множество приложений в различных областях человеческой деятельности,
а рассматриваемая в работе геометрическая ЗМОД непосредственно возникает в алгоритмах кластеризации. В этой работе для
каждого фиксированного $d\geq 2$ был предложен алгоритм сложности $O(n\cdot \log^{d-1}\,n)$ решения этой ЗМОД,
что при $d\geq 5$ улучшает ранее известную сложность $O\big(n\cdot (\log\,n + \log^{r_d}\,n\cdot \log\log\,n)\big)$,
где $r_d\in \{0,1,2,4\}$ при $d\in \{2,3,4,5\}$ и $r_d=d$ при $d\geq 6$, данной задачи.
\end{fulltext}

\end{document}